\documentclass[floatfix,letterpaper,english,reprint,secnumarabic,amssymb,nobibnotes,aps,prd,superscriptaddress]{revtex4-2}
\usepackage[T1]{fontenc}
\usepackage[latin9]{inputenc}
\usepackage{babel}
\usepackage{amsmath}
\usepackage{amssymb}
\usepackage{cancel}
\usepackage{graphicx}
\usepackage{wasysym}
\usepackage{physics}
\usepackage{multirow}
\usepackage{soul}
\usepackage[unicode=true,pdfusetitle,bookmarks=true,bookmarksnumbered=false,bookmarksopen=false,breaklinks=false,pdfborder={0 0 1},backref=false,colorlinks,linkcolor=blue,anchorcolor=blue,citecolor=blue]{hyperref}

\makeatletter

\pdfpageheight\paperheight
\pdfpagewidth\paperwidth

\makeatother

\begin{document}
\title{
    Direct lattice QCD calculation of the $\theta$-induced CP-violating pion-nucleon coupling
}

\author{Chuan-Yang Li}
\affiliation{State Key Laboratory of Nuclear Physics and Technology, Institute of Quantum Matter, South China Normal University, Guangzhou 510006, China}
\affiliation{Key Laboratory of Atomic and Subatomic Structure and Quantum Control~(MOE), Guangdong-Hong Kong Joint Laboratory of Quantum Matter, Guangzhou 510006, China}
\affiliation{Guangdong Basic Research Center of Excellence for Structure and Fundamental Interactions of Matter, Guangdong Provincial Key Laboratory of Nuclear Science, Guangzhou 510006, China}

\author{Jun Hua}
\affiliation{State Key Laboratory of Nuclear Physics and Technology, Institute of Quantum Matter, South China Normal University, Guangzhou 510006, China}
\affiliation{Guangdong Basic Research Center of Excellence for Structure and Fundamental Interactions of Matter, Guangdong Provincial Key Laboratory of Nuclear Science, Guangzhou 510006, China}

\author{Jian Liang}
\email{jianliang@scnu.edu.cn}
\affiliation{State Key Laboratory of Nuclear Physics and Technology, Institute of Quantum Matter, South China Normal University, Guangzhou 510006, China}
\affiliation{Guangdong Basic Research Center of Excellence for Structure and Fundamental Interactions of Matter, Guangdong Provincial Key Laboratory of Nuclear Science, Guangzhou 510006, China}

\author{Keh-Fei~Liu}
\affiliation{Department of Physics and Astronomy, University of Kentucky, Lexington,
KY 40506, USA}
\affiliation{Nuclear Science Division, Lawrence Berkeley National Laboratory, Berkeley, CA 94720, USA}

\author{Long-cheng Gui}
\affiliation{Department of Physics, Hunan Normal University,  Changsha 410081, China }
\affiliation{Synergetic Innovation
Center for Quantum Effects and Applications (SICQEA), Changsha 410081, China}
\affiliation{Key Laboratory of Low-Dimensional Quantum Structures and Quantum Control of Ministry of Education, Changsha 410081, China}

\author{Jun Shi}
\email{junshi@gdut.edu.cn}
\affiliation{Guangdong Provincial Key Laboratory of Sensing Physics and System Integration Applications,
School of Physics and Optoelectronic Engineering, Guangdong University of Technology,
Guangzhou, Guangdong 510006, China\\~\\
\includegraphics[scale=0.1]{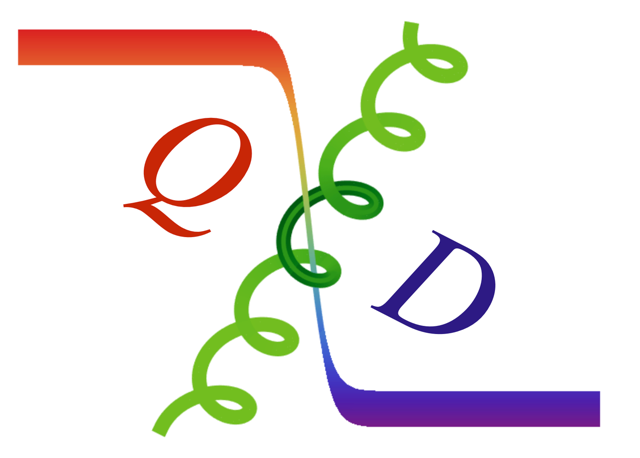}\\
{\rm($\chi$QCD Collaboration)}}
\author{Nan Wang}
\affiliation{State Key Laboratory of Nuclear Physics and Technology, Institute of Quantum Matter, South China Normal University, Guangzhou 510006, China}
\affiliation{Key Laboratory of Atomic and Subatomic Structure and Quantum Control~(MOE), Guangdong-Hong Kong Joint Laboratory of Quantum Matter, Guangzhou 510006, China}
\affiliation{Guangdong Basic Research Center of Excellence for Structure and Fundamental Interactions of Matter, Guangdong Provincial Key Laboratory of Nuclear Science, Guangzhou 510006, China}

\begin{abstract}
We present the first direct lattice QCD determination of the
$\theta$-induced CP-violating pion-nucleon-nucleon coupling
$\tilde g_{\pi NN}$.
Using overlap valence fermions on three $2+1$-flavor domain-wall
ensembles at a single lattice spacing, we calculate the forward proton
matrix element of the isovector pseudoscalar density in the $\theta$-vacuum 
to first order in $\bar\theta$.  The parity-mixing effect in the
external nucleon states is included in the extraction.  The
cluster-decomposition error-reduction method is used to improve the
statistical precision.  A simultaneous
extrapolation in the valence- and sea-pion masses, with model averaging
over 7 forms based on the Akaike information criterion, gives
$\tilde g_{\pi NN}=0.0306(56)(124)\,\bar\theta$ at the physical point.  The
first uncertainty includes the statistical and matrix-element-fit
systematic uncertainties, whereas the second reflects the spread among
the extrapolation forms.
Within these uncertainties, the result is consistent with the indirect
determination based on the strong neutron-proton mass splitting.  The
present precision is limited primarily by the extrapolation from the
relatively heavy sea-pion masses.
\end{abstract}

\maketitle

\section{Introduction}

CP violation is a necessary ingredient in explaining the
matter-antimatter asymmetry of the Universe.  However, the known sources of CP violation
in the Standard Model appear to be insufficient.  Searches for
permanent electric dipole moments (EDMs) therefore provide sensitive
probes of new CP-violating interactions~\cite{Chupp:2017rkp,Shindler:2021,Liu:2024kqy}.

At hadronic and nuclear scales, CP violation manifests itself through
nucleon EDMs and CP-violating interactions among nucleons.  The
long-range component of the CP-violating nuclear force arises from
one-pion exchange, with one CP-even and one CP-violating pion-nucleon
vertex.  The CP-violating pion-nucleon couplings are therefore
essential hadronic inputs for interpreting the EDMs of nucleons, light
nuclei, and diamagnetic atoms~\cite{Chupp:2017rkp,Shindler:2021}.  In
this work, we focus on this CP-violating pion-nucleon coupling induced by the QCD $\theta$ term.

We adopt the following convention
\begin{equation}
\mathcal{L}^{\cancel{CP}}_{\pi N}
=-\tilde g_{\pi NN}\,
 \bar N\boldsymbol{\tau}\cdot\boldsymbol{\pi}N+\cdots,
\label{eq:effective-coupling}
\end{equation}
where $N=(p,n)^T$. The coupling is related to the commonly used
dimensionful coupling $\bar g_0$ by
\begin{equation}
\tilde g_{\pi NN}=\frac{\bar g_0}{2F_\pi}.
\label{eq:gtilde-gbar0}
\end{equation}
Both $\bar g_0$ and $\tilde g_{\pi NN}$ are proportional to $\bar\theta$
at leading order. Equations~(\ref{eq:effective-coupling})
and~(\ref{eq:gtilde-gbar0}) fix the normalization used throughout this
work.

A direct lattice determination is challenging because the
$\theta$-induced signal is obtained from correlation functions weighted
by the global topological charge and is consequently very noisy.  Most
quantitative information has therefore been obtained indirectly.
Crewther et al. derived leading-order
$SU(3)$ relations between the CP-odd meson-baryon couplings and mass
splittings within the baryon octet~\cite{Crewther:1979pi}.  In
particular, $\bar g_0$ can be related at this order to the
$\Xi$-$\Sigma$ mass difference.  More recent chiral analyses showed
that a separate relation between $\bar g_0$ and the strong
neutron-proton mass splitting is preserved by loop corrections
through the orders studied, whereas the relation involving the
$\Xi$-$\Sigma$ masses receives large $SU(3)$-breaking corrections~\cite{deVries:2015gea,Seng:2016pfd}.  Unknown higher-order low-energy
constants can still generate corrections of about $10$-$20\%$~\cite{Seng:2016pfd}.

Combining the more robust neutron-proton matching relation with
lattice QCD determinations of the strong mass splitting gives
\begin{equation}
\tilde g_{\pi NN}=0.0155(20)(16)\,\bar\theta,
\label{eq:indirect-g}
\end{equation}
in the convention of Eq.~(\ref{eq:effective-coupling})~\cite{deVries:2015gea}.  Although this result uses lattice QCD input,
it does not directly evaluate the CP-odd pion-nucleon matrix element.
A direct calculation can therefore provide an independent test of the
chiral matching relation and of possible higher-order corrections.

In the present work, we determine $\tilde g_{\pi NN}$ directly from the
forward proton matrix element of the isovector pseudoscalar density.
The term linear in $\bar\theta$ is isolated with a
topological-charge-weighted three-point function.  To our knowledge,
this is the first direct lattice QCD calculation of the
$\theta$-induced coupling.  We employed the same gauge ensembles and
overlap valence action as in our previous nucleon EDM calculation~\cite{Liang:2023}.  The overlap definition of the topological charge is
consistent with the index theorem, while the cluster-decomposition
error-reduction method (CDER)~\cite{Liang:2023} is used to suppress the volume-enhanced
statistical noise.

The rest of this paper is organized as follows.  The $\bar\theta$
expansion, parity mixing, and the direct extraction of
$\tilde g_{\pi NN}$ are summarized in
Sec.~\ref{sec:formalism}.  The lattice setup, matrix-element analysis,
CDER study, and chiral extrapolation are described in
Sec.~\ref{sec:numerics}.  Section~\ref{sec:summary} discusses the main
uncertainties, compares the result with indirect determinations, and
summarizes our conclusions and outlook.

\section{Formalism}
\label{sec:formalism}

\subsection{\texorpdfstring{$\bar\theta$}{theta-bar} expansion and parity mixing}

The Euclidean QCD Lagrangian in the presence of the
$\theta$ term is
\begin{equation}
\mathcal{L}_{\bar\theta}(x)=\mathcal{L}_{\mathrm{QCD}}(x)-i\bar\theta q(x),
\qquad
Q=\int d^4x\,q(x),
\label{eq:theta-lagrangian}
\end{equation}
where $\bar{\theta}$ is the shifted $\theta$ which includes
the argument of the quark mass determinant, and 
\begin{equation}
q(x)=\frac{1}{32\pi^2}\epsilon_{\mu\nu\rho\sigma}
\operatorname{Tr}\left[G_{\mu\nu}(x)G_{\rho\sigma}(x)\right].
\label{eq:topological-density}
\end{equation}
The experimental bound on the neutron EDM requires $\bar\theta$ to be
very small~\cite{Abel:2020nEDM}.  Observables can therefore be expanded perturbatively
about the CP-even vacuum,
\begin{equation}
\langle\mathcal O\rangle_{\bar\theta}
=\langle\mathcal O\rangle_0
+i\bar\theta\langle\mathcal O Q\rangle_0+O(\bar\theta^2).
\label{eq:theta-expansion}
\end{equation}
CP symmetry at $\bar\theta=0$ implies $\langle Q\rangle_0=0$.  This
expansion allows the term linear in $\bar\theta$ to be evaluated on
ordinary CP-even gauge ensembles, avoiding simulations at nonzero real
$\bar\theta$.  Our Euclidean conventions and detail derivations of the correlation functions are
given in Ref.~\cite{Liang:2023}.

On a generic lattice, the topological charge can be evaluated from a
gluonic discretization of $G_{\mu\nu}\widetilde G_{\mu\nu}$.  Because
this operator is strongly affected by ultraviolet fluctuations, the
gauge fields are usually smoothed by gradient flow~\cite{Luscher:2010iy,Luscher:2011bx} before the charge is measured.  At
finite lattice spacing, the resulting charge can depend
on the chosen flow time and on the discretization of the gluonic
operator.  For a lattice formulation with exact chiral symmetry, an
alternative fermionic definition is available through the chiral Dirac
operator.  In our overlap fermion formulation, we use  the local charge from the overlap Dirac operator~\cite{Kikukawa:1998pd,Adams:1998eg,Fujikawa:1998if,Suzuki:1998yz}
\begin{equation}
q_t(x)=\frac{1}{2}\operatorname{Tr}
\left[\gamma_5D_{\mathrm{ov}}(x,x)\right],
\qquad Q=\sum_x q_t(x).
\label{eq:overlap-charge}
\end{equation}
where the trace runs over color and spin.  The spacetime sum of this
fermionic density satisfies the index theorem and requires no additional smoothing.  In Ref.~\cite{Liang:2023}, we compared this definition with the
gradient-flowed gluonic definition on the same ensembles.  The overlap
definition avoids the choice of a flow time and shows smaller discrete effects at the single lattice spacing used here.  We therefore adopt
it for the present calculation.

In addition to modifying correlation functions, the $\theta$ term
mixes opposite-parity components of the nucleon state.  To first order
in $\bar\theta$, the corresponding spinor can be written as
\begin{equation}
u_{\bar\theta}(p)=e^{i\alpha_1\bar\theta\gamma_5}u(p).
\label{eq:parity-mixing}
\end{equation}
The mixing angle $\alpha_1$ is determined from the ordinary and
$Q$-weighted proton two-point functions,
\begin{align}
G_p^{(2)}(t,\mathbf p)
={}&\sum_{\mathbf x}e^{-i\mathbf p\cdot\mathbf x}
\langle\chi_p(t,\mathbf x)\bar\chi_p(0)\rangle,
\label{eq:two-point}\\
G_p^{(2)Q}(t,\mathbf p)
={}&\sum_{\mathbf x}e^{-i\mathbf p\cdot\mathbf x}
\langle\chi_p(t,\mathbf x)Q\bar\chi_p(0)\rangle.
\label{eq:two-point-Q}
\end{align}
At zero momentum, their large-time ratio gives
\begin{equation}
\alpha_1=
\frac{\operatorname{Tr}\left[\gamma_5G_p^{(2)Q}(t,\mathbf 0)\right]}
{2\operatorname{Tr}\left[\Gamma_eG_p^{(2)}(t,\mathbf 0)\right]},
\qquad
\Gamma_e=\frac{1+\gamma_4}{2}.
\label{eq:alpha1-ratio}
\end{equation}
The traces in Eq.~(\ref{eq:alpha1-ratio}) are over spin indices. 

The external-state rotation in Eq.~(\ref{eq:parity-mixing}) must be
included in the extraction of a CP-odd matrix element.  As clarified
for lattice form-factor calculations in
Ref.~\cite{Abramczyk:2017oxr}, neglecting this rotation in an nEDM
calculation causes the CP-even Pauli form factor $F_2$ to contaminate
the CP-odd form factor $F_3$.  An analogous mixing occurs here: the
CP-even pseudoscalar charge $g_P$ contributes to the desired CP-odd
matrix element through a term proportional to $\alpha_1$, as discussed
in detail in the next subsection.

\subsection{Direct extraction of
\texorpdfstring{$\tilde g_{\pi NN}$}{the CP-violating pion-nucleon coupling}}

We take both the initial and final nucleons at zero spatial momentum,
so that the momentum transfer vanishes.  The inserted operator is the
Euclidean isovector pseudoscalar density
\begin{equation}
P^{u-d}(x)=\bar u(x)\gamma_5u(x)-\bar d(x)\gamma_5d(x).
\label{eq:pseudoscalar-density}
\end{equation}
For degenerate $u$ and $d$ quarks, disconnected contractions cancel in
this isovector combination.  Both $P^{u-d}$ and the topological charge
$Q$ are odd under parity, so their product $QP^{u-d}$ is a scalar.  Its
forward matrix element can therefore be extracted with the
unpolarized projector $\Gamma_e$, without introducing nonzero momentum transfers
or a polarized projector.

At zero momentum, we parameterize the proton matrix element in the
$\theta$-vacuum as
\begin{equation}
\langle p_{\bar\theta}(\mathbf 0)|P^{u-d}|p_{\bar\theta}(\mathbf 0)\rangle
=\bar u_{p,\bar\theta}(\mathbf 0)
\left[\gamma_5g_P+i\bar\theta\widetilde g'_P\right]
u_{p,\bar\theta}(\mathbf 0),
\label{eq:pseudoscalar-decomposition}
\end{equation}
Here $g_P$ and $\widetilde g'_P$ are the corresponding pseudoscalar
charges (couplings), defined by the form factors at zero momentum
transfer: $g_P\equiv G_P(q^2=0)$ and
$\widetilde g'_P\equiv\widetilde G'_P(q^2=0)$.  The prime indicates the
CP-odd scalar vertex before the parity rotation of the external
nucleon states is included.  The corresponding physical CP-odd charge
will be denoted by $\widetilde g_P$.  Because the calculation is
performed directly at $q^2=0$, no momentum-dependent fit is needed.

If the additional phase of the nucleon spinor were omitted, the
$\gamma_5g_P$ term would vanish under the unpolarized forward
projection at zero momentum, and only the direct CP-odd vertex
$\widetilde g'_P$ would contribute to the CP-odd charge.  In our
Euclidean convention, the phase in Eq.~(\ref{eq:parity-mixing}) enters
the adjoint spinor with the same sign,
$\bar u_{p,\bar\theta}=\bar u_p e^{i\alpha_1\bar\theta\gamma_5}$~\cite{Abramczyk:2017oxr}.
To display the effect of the two external-state rotations, we
substitute Eq.~(\ref{eq:parity-mixing}) into
Eq.~(\ref{eq:pseudoscalar-decomposition}) and expand each external
phase to first order,
\begin{equation}
\begin{aligned}
&~\langle p_{\bar\theta}|P^{u-d}|p_{\bar\theta}\rangle\\
\simeq&~\bar u_p
\left(1+i\alpha_1\bar\theta\gamma_5\right)
\left(\gamma_5g_P+i\bar\theta\widetilde g'_P\right)
\left(1+i\alpha_1\bar\theta\gamma_5\right)u_p\\
=&~\bar u_p\left[
\gamma_5g_P+i\bar\theta
\left(\widetilde g'_P+2\alpha_1g_P\right)
\right]u_p+O(\bar\theta^2).
\end{aligned}
\label{eq:pseudoscalar-phase-expansion}
\end{equation}
Equation~(\ref{eq:pseudoscalar-phase-expansion}) makes the
parity-mixing contribution explicit.  At $O(\bar\theta)$, the rotations of
the incoming and outgoing nucleon states each contribute
$\alpha_1g_P$, while terms involving both rotations are of
$O(\bar\theta^2)$ and are neglected.  The physical CP-odd charge is
therefore
$\widetilde g_P=\widetilde g'_P+2\alpha_1g_P$.  The same treatment was
used in our nEDM calculation~\cite{Liang:2023}.

On the lattice, this physical combination is isolated from the
pseudoscalar three-point function, which is expanded as
\begin{equation}
G^{(3),\bar\theta}_{P^{u-d}}
=G^{(3)}_{P^{u-d}}+i\bar\theta G^{(3)Q}_{P^{u-d}}+O(\bar\theta^2),
\label{eq:three-point-expansion}
\end{equation}
where the coefficient of the linear term is the $Q$-weighted
correlator
\begin{equation}
G^{(3)Q}_{P^{u-d}}(t_f,\tau)
=\sum_{\mathbf x,\mathbf y}
\langle\chi_p(x)Q P^{u-d}(y)\bar\chi_p(0)\rangle,
\label{eq:three-point-Q}
\end{equation}
with $x=(t_f,\mathbf x)$ and $y=(\tau,\mathbf y)$.  In the ground-state
limit, inserting the nucleon ground state gives the same three-factor
structure as in Eq.~(\ref{eq:pseudoscalar-phase-expansion}),
\begin{equation}
\begin{aligned}
G^{(3),\bar\theta}_{P^{u-d}}
\propto{}& e^{-m_N t_f}
\left(\Gamma_e+i\bar\theta\alpha_1\gamma_5\right)
\left(\gamma_5g_P+i\bar\theta\widetilde g'_P\right)\\
&\times\left(\Gamma_e+i\bar\theta\alpha_1\gamma_5\right)\\
={}&i\bar\theta e^{-m_Nt_f}\,
\Gamma_e\left(\widetilde g'_P+2\alpha_1g_P\right)\Gamma_e
+O(\bar\theta^2),
\end{aligned}
\label{eq:three-point-ground-state}
\end{equation}
where the CP-even $g_P$ term $G_{P^{u-d}}^{(3)}$ has been omitted in the second line
because it vanishes under the unpolarized trace used below,
$\operatorname{Tr}[\Gamma_e\gamma_5\Gamma_e]=0$.
Comparing the coefficient of $i\bar\theta$ with
Eq.~(\ref{eq:three-point-expansion}) gives
\begin{equation}
\begin{aligned}
G^{(3)Q}_{P^{u-d}}
&\propto e^{-m_Nt_f}\,
\Gamma_e\left(\widetilde g'_P+2\alpha_1g_P\right)\Gamma_e\\
&=e^{-m_Nt_f}\widetilde g_P\Gamma_e.
\end{aligned}
\label{eq:three-point-Q-ground-state}
\end{equation}
Equation~(\ref{eq:three-point-Q-ground-state}) shows that the
$Q$-weighted three-point function directly isolates the complete
CP-odd charge $\widetilde g_P$, rather than the vertex contribution
$\widetilde g'_P$ alone.

Taking the ratio to the ordinary proton two-point function cancels the
overlap factors and the common Euclidean-time dependence, and yields
\begin{equation}
\begin{aligned}
R^{Q}_{P^{u-d}}(t_f,\tau)
&=\frac{\operatorname{Tr}
\left[\Gamma_eG^{(3)Q}_{P^{u-d}}(t_f,\tau)\right]}
{\operatorname{Tr}\left[\Gamma_eG_p^{(2)}(t_f,\mathbf 0)\right]}\\
&\xrightarrow{\tau\gg a,\,t_f-\tau\gg a}
2\alpha_1g_P+\widetilde g'_P
\equiv\widetilde g_P.
\end{aligned}
\label{eq:ratio-GP}
\end{equation}
Compared with our previous nEDM calculation, the present extraction
requires neither an extrapolation to zero momentum transfer nor
separate calculations of the mixing angle $\alpha_1$ and the CP-even
charge $g_P$~\cite{Liang:2023}.  Instead, the ratio in
Eq.~(\ref{eq:ratio-GP}) directly determines the full physical
combination $2\alpha_1g_P+\widetilde g'_P$, reducing both the
computational cost and the uncertainties from these additional
calculations.

The CP-odd pion-nucleon coupling is obtained
from the physical pseudoscalar form factor by factoring out the pion
pole,
\begin{equation}
\begin{aligned}
 m_q\widetilde G_P(q^2)
&=F_\pi\frac{m_\pi^2}{m_\pi^2-q^2}
\frac{\tilde g_{\pi NN}}{\bar\theta},\\
\frac{\tilde g_{\pi NN}}{\bar\theta}
&=\frac{m_q}{F_\pi}\widetilde g_P
=\frac{m_q}{F_\pi}
\left(2\alpha_1g_P+\widetilde g'_P\right),
\end{aligned}
\label{eq:pion-pole}
\end{equation}
where the pion-pole normalization is fixed by the nonsinglet PCAC
relation,
$m_q\langle 0|P^{u-d}|\pi^0\rangle=m_\pi^2F_\pi$.
The first line of Eq.~(\ref{eq:pion-pole}) reduces directly to the
second at $q^2=0$.  Here $m_q$ is the degenerate bare overlap
valence-quark mass.  For the nonsinglet pseudoscalar density, the exact
chiral symmetry of the overlap action gives $Z_mZ_P=1$, so the product
$m_q\widetilde G_P(q^2)$ is renormalization-group invariant.  No
separate quark-mass or pseudoscalar-density renormalization is therefore
required, which is another advantage of using the overlap formulation.

We use the physical value $F_\pi=92.2\,\mathrm{MeV}$, consistent with
the normalization adopted in the indirect chiral determination of
Ref.~\cite{deVries:2015gea}.

\begin{table*}[!t]
\caption{Ensemble labels, sea-pion masses, valence-pion masses,
corresponding CDER cutoffs, nucleon masses, and numbers of gauge
configurations.  Within each row, the CDER cutoff values are listed in
the same order as the valence-pion masses.  The ensemble parameters follow
Refs.~\cite{Aoki:2010dy,Liang:2023}.}
\label{tab:ensembles}
\begin{ruledtabular}
\begin{tabular}{lccccc}
Ensemble & $m_{\pi,\mathrm{sea}}$ (MeV)
& $m_{\pi,\mathrm{val}}$ (MeV) & CDER cutoff $R/a$ & $m_N$ (GeV) & $N_{\mathrm{cfg}}$ \\
\hline
24I005 & 339 & 282, 321, 348, 389 & 17, 17, 17, 17 & 1.14 & 805 \\
24I010 & 432 & 426, 519, 600      & 23, 23, 23     & 1.25 & 508 \\
24I020 & 560 & 432, 525, 606      & 17, 16, 14     & 1.30 & 552 \\
\end{tabular}
\end{ruledtabular}
\end{table*}

\section{Numerical details}
\label{sec:numerics}

\subsection{Lattice setup}

We use the same lattice setup as in our nucleon EDM calculation~\cite{Liang:2023},
including the gauge ensembles, valence quark masses, source construction,
smearing, and variance-reduction methods. The difference is that the electromagnetic
current is replaced by the isovector pseudoscalar density, and both the
initial and final nucleons are restricted to zero spatial momentum.

The calculation employs three $2+1$-flavor RBC/UKQCD domain-wall
ensembles~\cite{Aoki:2010dy}, all with volume $24^3\times64$ and lattice
spacing $a=0.1105(3)\,\mathrm{fm}$. The Iwasaki
gauge action and domain-wall fermions are used in the sea sector, while the valence quarks are
described by the overlap action on HYP-smeared gauge links~\cite{Li:2010pw}.
For each sea-pion mass, we calculate at multiple valence-pion masses.
This partially quenched setup helps disentangle the sea- and
valence-pion-mass dependence of $\tilde g_{\pi NN}$ and provides more
data to constrain the chiral extrapolation.  The ensemble parameters
are summarized in Table~\ref{tab:ensembles}; the selection of the CDER
cutoffs $R/a$ is discussed in Sec.~\ref{sec:cder}.

The massive overlap propagator is constructed from the chirally
symmetric operator $D_c$ as
\begin{subequations}
\label{eq:overlap-effective-propagator}
\begin{align}
G(m_q)&=(D_c+m_q)^{-1},
\label{eq:overlap-propagator}\\
D_c&=\frac{D_{\mathrm{ov}}}{1-D_{\mathrm{ov}}/2}.
\label{eq:chirally-symmetric-operator}
\end{align}
\end{subequations}
The operator $D_c$ satisfies the continuum anticommutation relation
$\{D_c,\gamma_5\}=0$, which ensures exact valence chiral symmetry at
nonzero lattice spacing and the relation $Z_mZ_P=1$ used
above~\cite{Neuberger:1997fp,Li:2010pw}.
Propagators for all valence masses are obtained with a multimass
inversion.

Nucleon propagators are generated from $2^3$ spatial grid sources with
$Z_3$ noise~\cite{Dong:1993pk}, with the grid origin randomly shifted on each gauge
configuration.  Sources at $t_{\mathrm{src}}/a=0$ and 32 are included
in the same inversion, and Gaussian smearing is applied to the nucleon
interpolating fields.  Low-mode substitution (LMS) reconstructs the
low-mode contribution from the eigenvectors separately for each grid
site.  When the nucleon correlation functions are assembled from the
grid-source quark propagators, this removes mixed-site noise from the
low-mode sector~\cite{Li:2010pw}.

The three-point functions are calculated with the stochastic-sandwich
method and also LMS~\cite{Yang:2015uis,Wang:2021vqy}, allowing multiple sink interpolating
operators and polarizations without separate sequential inversions.
We use three source-sink time separations,
$t_{\mathrm{sep}}/a\in\{6,8,10\}$. We use 8 independent source-noise
vectors and 16 sink-noise vectors at each separation.  The resulting
increase in statistics is essential for resolving the weak CP-odd
signal in the present calculation.

\begin{figure}[t]
    \centering
    \includegraphics[page=1,width=0.82\linewidth]{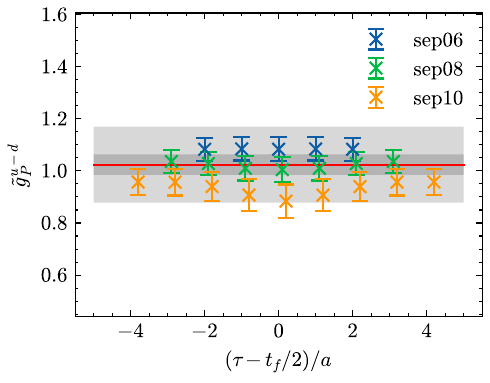}
    \par\vspace{0.1em}
    \includegraphics[page=2,width=0.82\linewidth]{fit_ME.pdf}
    \par\vspace{0.1em}
    \includegraphics[page=3,width=0.82\linewidth]{fit_ME.pdf}
    \caption{Representative constant fits to the CP-odd
    pseudoscalar charge.  From top to bottom, the sea and valence pion
    masses are $(339,321)$, $(432,519)$, and $(560,525)$, respectively,
    in MeV.  The data at $t_{\mathrm{sep}}/a\in\{6,8,10\}$ are shown.  The dark
    band denotes the statistical uncertainty of the constant fit,
    while the light band denotes the total uncertainty including
    the excited-state systematic uncertainty.}
    \label{fig:me-fit-pages-1-3}
\end{figure}

\subsection{Matrix-element fits}

For the ratio in Eq.~(\ref{eq:ratio-GP}), the individual
flavor-diagonal $u$ and $d$ contributions show visible
source-sink-separation dependence, but this dependence largely cancels
in the $u-d$ combination.  Figure~\ref{fig:me-fit-pages-1-3}
illustrates this cancellation and the absence of statistically
significant residual dependence.  Therefore, unlike in our previous
nucleon EDM analysis, where two-state fits were
used~\cite{Liang:2023}, we perform a combined constant fit to the ratios at
$t_{\mathrm{sep}}/a\in\{6,8,10\}$.  We retain all insertion times except
those that coincide with the source or sink.  The resulting fit bands
are included in the same figure.  Statistical uncertainties and the
covariance matrix are estimated by jackknife resampling.  The fits are
correlated, with an SVD cutoff of $10^{-4}$ used to regularize the
covariance-matrix inversion.  The fitted charges are stable under
variations of this cutoff.  For all mass combinations, the correlated
fits yield acceptable $\chi^2/\mathrm{dof}$ at both the selected CDER
cutoffs and larger cutoff values.

The lack of statistically resolved separation dependence does not
imply that excited-state effects are absent; the effects are
obscured by the statistical uncertainties.  We estimate the
corresponding systematic uncertainty using the largest separation with
a statistically significant signal, denoted by
$t_{\mathrm{sep}}^{\mathrm{ref}}$.  The systematic uncertainty is
defined as
\begin{equation}
\delta_{\mathrm{ES}}=
\max\!\left(
\left|\widetilde g_P^{\mathrm{const}}
-\widetilde g_{P,\mathrm{ref}}^{\mathrm{const}}\right|,
\left|\widetilde g_P^{\mathrm{const}}
-\widetilde g_{P,\mathrm{ref}}^{\mathrm{mid}}\right|
\right).
\label{eq:excited-state-systematic}
\end{equation}
Here $\widetilde g_P^{\mathrm{const}}$ is obtained from the combined
constant fit over all three source-sink separations,
$\widetilde g_{P,\mathrm{ref}}^{\mathrm{const}}$ from a constant fit
using only the data at $t_{\mathrm{sep}}^{\mathrm{ref}}$, and
$\widetilde g_{P,\mathrm{ref}}^{\mathrm{mid}}$ from the ratio at the
central insertion time of the same separation.  Thus,
$\delta_{\mathrm{ES}}$ is the larger of the shifts from the combined-fit
result to these two reference-separation estimates.  We use
$t_{\mathrm{sep}}^{\mathrm{ref}}/a=10$ whenever
the signal is statistically significant and otherwise use
$t_{\mathrm{sep}}^{\mathrm{ref}}/a=8$.  In
Fig.~\ref{fig:me-fit-pages-1-3}, the dark bands show the statistical
uncertainty of the combined fit, and the light bands show the total
uncertainty,
\begin{equation}
\delta_{\mathrm{tot}}=
\sqrt{\delta_{\mathrm{stat}}^2+\delta_{\mathrm{ES}}^2}.
\label{eq:matrix-element-total-uncertainty}
\end{equation}
We use this total uncertainty for each input point in the subsequent
chiral analysis.

\subsection{CDER analysis}
\label{sec:cder}

\begin{figure}[!h]
    \centering
    \includegraphics[page=4,width=0.82\linewidth]{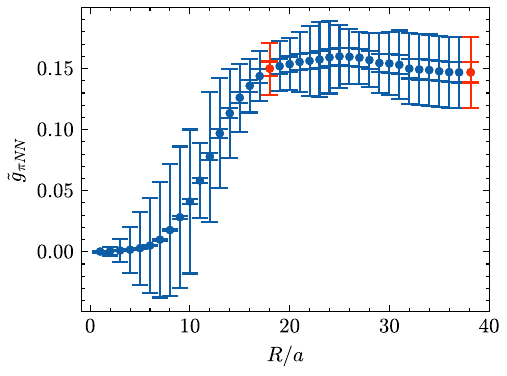}
    \par\vspace{0.1em}
    \includegraphics[page=5,width=0.82\linewidth]{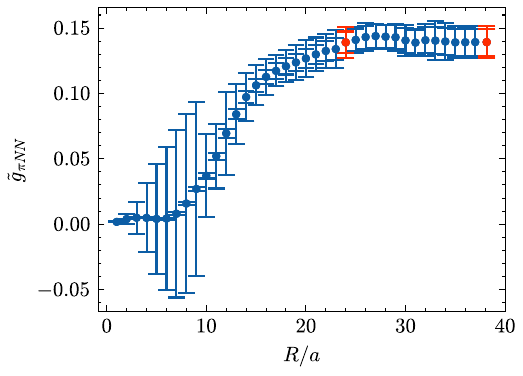}
    \par\vspace{0.1em}
    \includegraphics[page=6,width=0.82\linewidth]{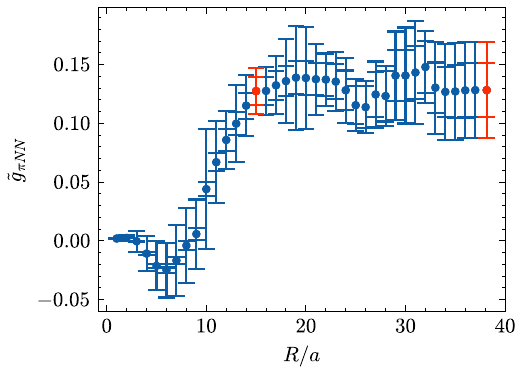}
    \caption{Dependence of $\tilde g_{\pi NN}$ on the four-dimensional
    CDER cutoff $R$ for three representative mass combinations.  From
    top to bottom, the sea and valence pion masses are $(339,321)$,
    $(432,519)$, and $(560,606)$, respectively, in MeV.  In each panel,
    the first red point marks the CDER cutoff selected for the preferred
    analysis, while the second is the maximum cutoff and corresponds to
    the untruncated full-volume sum used to assess the effect of the
    CDER cutoff.}
    \label{fig:me-fit-pages-4-6}
\end{figure}

Inserting the global topological charge generates a variance that grows
with the lattice volume.  To suppress this volume-enhanced noise, we
apply the cluster-decomposition error-reduction (CDER)
method~\cite{Liu:2017man}.  In the three-point function, the global
charge in Eq.~(\ref{eq:three-point-Q}) is replaced by a sum of the local
topological charge density within a four-dimensional CDER cutoff $R$
centered on the pseudoscalar insertion,
\begin{equation}
Q P^{u-d}(y)\longrightarrow
\sum_{|r|<R}q_t(y+r)P^{u-d}(y).
\label{eq:cder}
\end{equation}
Here $|r|$ is the shortest four-dimensional Euclidean distance between
the two sites after accounting for the periodic boundary conditions.
Cluster decomposition implies that the Euclidean correlation decreases exponentially.
Once the cutoff $R$ exceeds the relevant correlation length,
extending the sum farther primarily accumulates
statistical noise, but not signal.

For each CDER cutoff, we define the four-dimensional mask
$M_R(r)=\Theta(R-|r|)$, where $r$ is the displacement from the origin,
and the corresponding local sum of the
topological charge,
\begin{equation}
Q_M(R,x)=\sum_r M_R(r)q_t(x+r).
\label{eq:cder-local-charge}
\end{equation}
Thus, $Q_M(R,x)$ is the topological charge summed within a
four-dimensional distance $R$ of the point $x$.  We compute this
convolution by Fourier transforming $q_t$ and $M_R$, multiplying their
Fourier components, and applying an inverse Fourier transform,
\begin{equation}
Q_M(R,x)=\mathcal F^{-1}
\left[\widetilde M_R(k)\widetilde q_t(k)\right](x).
\label{eq:cder-fft}
\end{equation}
The $R$ dependent three-point function $G^{(3)Q}_{P^{u-d}}$ is then obtained directly from
$\langle\sum_{\mathbf x}C_3(x)Q_M(R,x)\rangle$, where $C_3(x)$ denotes the
three-point propagator function with the pseudoscalar insertion at $x$.  This procedure reduces the computational scaling of
constructing $Q_M(R,x)$ for all centers $x$ from quadratic in the
lattice volume $V$ to approximately $V\log V$~\cite{Liu:2017man}.

To study the cutoff dependence, we sample $R$ in steps of one lattice spacing,
$R/a=1,2,\ldots$. Because the exact
distances between lattice sites take the form $\sqrt{n}\,a$, nearby
separations are radially averaged within four-dimensional shells of
width approximately $a$.  At large $R$, each shell contains many
displacement vectors, and the cumulative results at neighboring CDER
cutoffs become almost fully correlated. We therefore do not perform a constant fit to
the points beyond the selected CDER cutoff.

Figure~\ref{fig:me-fit-pages-4-6} shows that the cutoff dependence
reaches a saturation region for all mass combinations.  The selected
optimal cutoff $R$ in each panel is marked by the first red point,
where the signal has saturated while the gain in statistical precision
is retained.  Ordered by increasing valence pion mass, the selected
CDER cutoffs are
$R/a=(17,17,17,17)$ on 24I005, $(23,23,23)$ on 24I010, and
$(17,16,14)$ on 24I020, as listed in Table~\ref{tab:ensembles}.  Results
at these cutoffs define our preferred analysis.  For comparison, the
second red point in each panel corresponds to the largest possible
cutoff, which includes the full four-dimensional volume, and is used
below to estimate the effect of the CDER cutoff.

\subsection{Chiral extrapolation and final result}

Because the partially quenched data exhibit distinct valence- and
sea-pion-mass dependence, we extrapolate $\tilde g_{\pi NN}/\bar\theta$
simultaneously in
$m_{\pi,\mathrm{val}}^2$ and $m_{\pi,\mathrm{sea}}^2$.  Each input
point is assigned the total matrix-element uncertainty defined in
Eq.~(\ref{eq:matrix-element-total-uncertainty}), which includes both
the statistical and excited-state systematic contributions.

\begin{figure}[!h]
    \centering
    \includegraphics[page=1,width=0.80\linewidth]{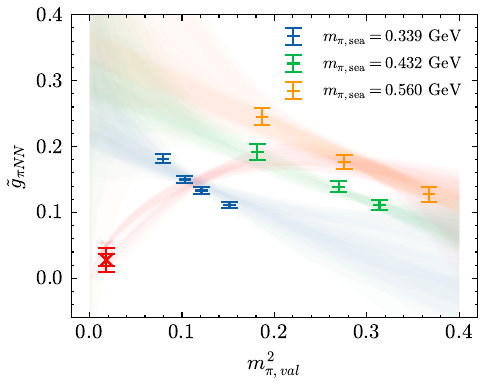}
    \par\vspace{0.2em}
    \includegraphics[page=2,width=0.80\linewidth]{fit_extrapolation.pdf}
    \caption{Simultaneous sea- and valence-pion-mass extrapolations of $\tilde g_{\pi NN}/\bar\theta$.  The upper panel uses
    the selected mass-dependent CDER cutoffs in
    Table~\ref{tab:ensembles}, whereas the lower panel uses the largest
    cutoff and corresponds to the untruncated full-volume analysis.
    The blue, green, and orange bands represent the 7 extrapolation
    forms at fixed sea-pion masses
    $m_{\pi,\mathrm{sea}}=0.339$, $0.432$, and
    $0.560\,\mathrm{GeV}$, respectively, as functions of
    $m_{\pi,\mathrm{val}}^2$, while the red bands give the corresponding
    unitary extrapolations along
    $m_{\pi,\mathrm{sea}}=m_{\pi,\mathrm{val}}$.  The red point in each
    panel denotes the AIC-weighted result at the physical pion mass;
    the shorter error bar represents $\sigma_{\mathrm{data}}$, while
    the longer one gives the total uncertainty including both
    $\sigma_{\mathrm{data}}$ and $\sigma_{\mathrm{form}}$.}
    \label{fig:extrapolation-results}
\end{figure}

Since no dedicated partially quenched chiral expression is currently
available for this observable, and the present data do not favor a
unique empirical parametrization, we consider the following seven
candidate forms for $\tilde g_{\pi NN}/\bar\theta$, inspired by the
partially quenched chiral perturbation theory expression for the
nucleon EDM in Ref.~\cite{OConnell:2005}:
\begingroup
\setlength{\jot}{1pt}
\begin{subequations}
\label{eq:chiral-ansatze}
\begin{align}
\mathcal{F}_0={}&a m_{\pi,v}^2+b m_{\pi,s}^2,
\label{eq:chiral-f0}\\
\mathcal{F}_1={}&a m_{\pi,v}^2+b m_{\pi,s}^2+c m_{\pi,s}^4,
\label{eq:chiral-f1}\\
\mathcal{F}_2={}&a m_{\pi,v}^2+b m_{\pi,v}^4+c m_{\pi,s}^2,
\label{eq:chiral-f2}\\
\mathcal{F}_3={}&a m_{\pi,v}^2+b m_{\pi,v}^4+c m_{\pi,s}^2+d m_{\pi,s}^4,
\label{eq:chiral-f3}\\
\mathcal{F}_4={}&a m_{\pi,v}^2+b m_{\pi,v}^4+c m_{\pi,s}^2+d m_{\pi,s}^2\ell_v,
\label{eq:chiral-f4}\\
\mathcal{F}_5={}&a m_{\pi,v}^2+b m_{\pi,s}^2+c m_{\pi,s}^4+d m_{\pi,s}^2\ell_v,
\label{eq:chiral-f5}\\
\mathcal{F}_6={}&a m_{\pi,v}^2+b m_{\pi,s}^2+c m_{\pi,s}^2\ell_v+d m_{\pi,s}^2\ell_s.
\label{eq:chiral-f6}
\end{align}
\end{subequations}
\endgroup
Here $a$, $b$, $c$, and $d$ are fit parameters.  For compact notation
in Eq.~(\ref{eq:chiral-ansatze}), we define
$m_{\pi,v}=m_{\pi,\mathrm{val}}$,
$m_{\pi,s}=m_{\pi,\mathrm{sea}}$,
$\ell_v=\log(m_{\pi,v}^2/\Lambda^2)$, and
$\ell_s=\log(m_{\pi,s}^2/\Lambda^2)$,
and we use $\Lambda=1\,\mathrm{GeV}$ numerically.
Changing $\Lambda$ only shifts the logarithms by a constant; in each
ansatz, this shift can be absorbed into the coefficients of the
polynomial mass terms, leaving the functional form unchanged.
A further common feature of the seven ansatzes is that they vanish in
the chiral limit along the unitary trajectory
$m_{\pi,\mathrm{val}}=m_{\pi,\mathrm{sea}}$.  This behavior is
consistent with the lattice chiral fermion formulations employed in this study,
which preserve the expected chiral limit even at nonzero lattice spacing.
For each form, the
physical-point prediction is obtained by setting
$m_{\pi,\mathrm{val}}=m_{\pi,\mathrm{sea}}=m_{\pi,\mathrm{phys}}$, where
$m_{\pi,\mathrm{phys}}=135\,\mathrm{MeV}$.

Rather than selecting a single preferred parametrization, we combine the seven
fits using the standard Akaike information criterion
(AIC)~\cite{Akaike:1974,Borsanyi:2014jba}.  A fit $i$ with chi-squared
$\chi_i^2$ and $k_i$ free parameters is assigned a weight
\begin{align}
w_i={}&\frac{\exp(-\mathrm{AIC}_i/2)}
{\sum_j\exp(-\mathrm{AIC}_j/2)},
\label{eq:aic-weight}\\
\mathrm{AIC}_i={}&\chi_i^2+2k_i.
\label{eq:aic}
\end{align}
Denoting the physical-point value of $\tilde g_{\pi NN}/\bar\theta$ from
fit $i$ by $x_i$ and its uncertainty propagated from the total input
uncertainties by $\sigma_i^{\mathrm{data}}$, we define the corresponding
model average and its two uncertainty components as
\begin{align}
\bar x={}&\sum_i w_i x_i,
\label{eq:aic-average}\\
(\sigma_{\mathrm{data}})^2={}&
\sum_i w_i(\sigma_i^{\mathrm{data}})^2,
\label{eq:aic-data-uncertainty}\\
(\sigma_{\mathrm{form}})^2={}&
\sum_i w_i(x_i-\bar x)^2.
\label{eq:aic-form-uncertainty}
\end{align}
Here $\sigma_{\mathrm{data}}$ incorporates the statistical and
excited-state systematic uncertainties of the matrix elements through
the extrapolation, whereas $\sigma_{\mathrm{form}}$ quantifies the
AIC-weighted spread among the candidate forms.  The 7 individual fits
are shown in Fig.~\ref{fig:extrapolation-results}: the blue, green, and
orange bands describe their valence-pion-mass dependence at the three
simulated sea-pion masses, whereas the red bands trace their unitary
continuations toward the physical point.  The red point in each panel
marks the corresponding AIC-weighted physical-point result.  The
shorter error bar represents $\sigma_{\mathrm{data}}$, whereas the
longer one gives the total uncertainty including both
$\sigma_{\mathrm{data}}$ and $\sigma_{\mathrm{form}}$.

Applying this procedure to the data at the selected optimal CDER
cutoffs and restoring the explicit factor of $\bar\theta$, we obtain
\begin{equation}
\tilde g_{\pi NN}=0.0306(56)(124)\,\bar\theta,
\label{eq:final-result}
\end{equation}
where the first uncertainty is $\sigma_{\mathrm{data}}$ and the second
is $\sigma_{\mathrm{form}}$.  As a check of the CDER-cutoff dependence,
we repeat the complete analysis using the full-volume sums and find the
consistent, though less precise, result
\begin{equation}
\tilde g_{\pi NN}^{\mathrm{full\ volume}}=0.0305(87)(140)\,\bar\theta.
\label{eq:full-volume-result}
\end{equation}
The agreement between Eqs.~(\ref{eq:final-result}) and
(\ref{eq:full-volume-result}) indicates that the selected CDER cutoffs
improve the precision without affecting the physical result.

\section{Discussion and Summary}
\label{sec:summary}

Our preferred direct result in Eq.~(\ref{eq:final-result}) carries two
distinct uncertainties.  The first incorporates the statistical and
excited-state systematic uncertainties of the matrix elements.  The
second is the AIC-weighted spread of the physical-point values obtained
from the 7 extrapolation forms in Eq.~(\ref{eq:chiral-ansatze}).
Although each form describes the data over the simulated mass range,
they yield appreciably different results when extrapolated to the
physical point.  This spread directly measures the fit-form dependence
and, more broadly, reflects the challenges of extrapolating from the
relatively heavy sea-pion masses used here to the physical point,
particularly in the absence of a dedicated partially quenched chiral
perturbation theory expression for this observable.  The uncertainty
can therefore be
reduced either by adding ensembles with lighter sea-pion masses or by
deriving a more strongly constrained partially quenched chiral form;
calculations pursuing the former improvement are in progress.

Within these uncertainties, the direct result is consistent with the
indirect determination from the strong neutron-proton mass splitting
in Eq.~(\ref{eq:indirect-g})~\cite{deVries:2015gea}.  A remaining
limitation is that all three ensembles share a single lattice spacing;
the quoted uncertainties therefore do not include discretization
effects, whose quantification will require calculations at additional
lattice spacings.

Our partially quenched data reveal distinct valence- and
unitary-pion-mass dependences. As shown in
Fig.~\ref{fig:extrapolation-results}, when the sea-pion mass is held
fixed, the magnitude of $\tilde g_{\pi NN}$ increases as the
valence-pion mass is lowered; along the unitary trajectory
$m_{\pi,\mathrm{val}}=m_{\pi,\mathrm{sea}}$, however, it decreases
toward the chiral limit, where the $\theta$-induced strong
CP-violating effect vanishes.  This contrast is expected because
varying only the valence-pion mass does not approach the unitary chiral
limit.  Although the same qualitative behavior was observed in our
previous nEDM calculation~\cite{Liang:2023}, the stronger signal for
$\tilde g_{\pi NN}$ makes it considerably clearer in the present data.

In our previous nEDM calculation, a dedicated partially quenched
chiral expression provided a tighter constraint on the mass dependence
and resulted in a substantially smaller extrapolation-form
uncertainty than in the present analysis~\cite{OConnell:2005,Liang:2023}.
That calculation gave
\begin{equation}
d_n=-0.00148(14)(31)\,\bar\theta\,e\,\mathrm{fm}.
\label{eq:previous-nedm}
\end{equation}
For a qualitative cross-check, we retain only the long-distance
pion-loop logarithm derived in Ref.~\cite{Crewther:1979pi}, which in our
convention relates the neutron EDM to $\tilde g_{\pi NN}$ through
\begin{equation}
\frac{d_n}{e}
\simeq-\frac{g_{\pi NN}\tilde g_{\pi NN}}
{4\pi^2M_N}\log\frac{M_N}{m_\pi},
\label{eq:nedm-chiral-relation}
\end{equation}
where the minus sign arises from translating the convention of that
reference to the one adopted in Eq.~(\ref{eq:effective-coupling}).
Using
$g_{\pi NN}=13.2$~\cite{Reinert:2020mcu} and the nEDM result in
Eq.~(\ref{eq:previous-nedm}) gives the rough estimate
$\tilde g_{\pi NN}^{(d_n)}\simeq0.0109\,\bar\theta$.
Numerically, this estimate is compatible with both the mass-splitting
determination in Eq.~(\ref{eq:indirect-g}) and our direct result.  This
estimate should not be interpreted as a controlled determination,
because both analytic contributions and higher-order chiral corrections
are omitted.  Moreover, the nEDM determination and the direct
calculation are subject to different systematic uncertainties, such as physical pion mass, continuum and infinite volume extrapolations. The
direct calculation is therefore conceptually distinct and remains our
primary result.

In summary, we have directly determined the $\theta$-induced
CP-violating pion-nucleon coupling using overlap valence fermions on
three domain-wall ensembles.  The analysis accounts for the
parity-mixed external-state contribution and employs CDER to improve
the statistical precision.  Combining the 7 simultaneous
valence- and sea-pion-mass extrapolations through AIC model averaging,
we obtain
\begin{equation*}
\tilde g_{\pi NN}=0.0306(56)(124)\,\bar\theta,
\end{equation*}
with the present precision limited primarily by the chiral
extrapolation.  Calculations at lighter pion masses are under way to
reduce the extrapolation range and sharpen the direct determination,
while simulations at additional lattice spacings will ultimately be
needed to quantify discretization effects.

\section*{Acknowledgments}
The numerical calculation of this work are performed on the NERSC supercomputer system Perlmutter and Southern Nuclear Science Computing Center (SNSC) of South China Normal University.
This work is supported by the National Natural Science Foundation of China (NSFC) under Grants No.\ 12575085, No.\ 12105108.
This work is supported in part by the U.S. Department of Energy, Office of Science, Office of Nuclear Physics, under Grant No. DE-SC0013065. The authors acknowledge partial support by the U.S. Department of Energy, Office of Science, Office of Nuclear Physics under the umbrella of the Quark-Gluon Tomography (QGT) Topical Collaboration with Award No. DE-SC0023646. This research used resources of the National Energy Research Scientific Computing Center (NERSC), a U.S. Department of Energy Office of Science User Facility located at Lawrence Berkeley National Laboratory, operated under Contract No. DE-AC02-05CH11231. This work used computational resources on Stampede3 and Frontera at the Texas Advanced Computing Center (TACC) at The University of Texas at Austin. Stampede3 is supported by the National Science Foundation under award OAC-2311290, and Frontera is supported by the National Science Foundation under award OAC-1818253. We also acknowledge the facilities of the USQCD collaboration used for this research in part, which are funded by the Office of Science of the U.S. Department of Energy.

\bibliography{main}

\end{document}